# Relative Focal Mechanism Inversion Using Relative Polarities and S/P Double Ratios

Miao Zhang[1*]

[1]Department of Earth and Environmental Sciences, Dalhousie University, Halifax, Nova Scotia, Canada

[*]Corresponding author: miao.zhang@dal.ca

**Abstract**

Focal mechanisms of small earthquakes are critical for characterizing faults and regional stress. P-wave polarities and S/P amplitude ratios (e.g., in HASH) are widely used to determine focal mechanisms for small earthquakes, but first-motion picking can be difficult for emergent onsets, and S/P ratios are often highly inconsistent because of imperfect velocity models and unknown site responses. Here, I present FocMecDR, a relative focal-mechanism inversion method that uses relative polarities and double ratios of S/P amplitudes. Analogous to double-difference earthquake location, FocMecDR uses a reference event with a known focal mechanism to solve focal mechanisms for nearby earthquakes by minimizing the misfit between observed and theoretical S/P amplitude double ratios while enforcing cross-correlation-based relative polarities. Extensive synthetic tests demonstrate the effectiveness and accuracy of the method. Field verification using an antisimilar earthquake pair and aftershocks from the northwestern end of the 2019 Ridgecrest rupture zone further confirms its reliability. Application to the foreshocks preceding the 2019 Ridgecrest mainshock reveals a highly uniform stress field during nucleation, suggesting a possible framework for evaluating fault-zone stress and rupture readiness before large earthquakes.

**1. Introduction**

Focal mechanisms provide critical information about earthquake source processes and fault geometry (Aki and Richards, 2002). Beyond characterizing the physical rupture, they offer essential constraints on the regional stress field (Vavryčuk, 2014), thereby contributing to tectonic interpretations and seismic hazard assessment. Although full moment tensor solutions contain more comprehensive information (Jost and Herrmann, 1989), including the ability to distinguish explosions from earthquakes (Cesca et al., 2017) and to resolve non–double-couple components related to specific settings such as fluid environments (Wang et al., 2018), their reliable estimation requires dense station coverage, high-quality data, and accurate velocity models. In many regional studies, particularly where station coverage is limited, double-couple solutions remain more robust, computationally feasible, and sufficient for both research and industry applications. Currently two primary methodological approaches are commonly used for double-couple-based focal mechanism inversion: waveform fitting methods and first-motion/amplitude-ratio methods.

Waveform fitting has been widely applied in focal mechanism inversion. These methods involve forward modeling of synthetic waveforms and minimizing the misfit between synthetic and observed data. Their accuracy depends strongly on the quality of the seismic velocity model. In small-scale industrial studies (e.g., with well-log constraints), velocity structures can be well constrained (e.g., Li et al., 2011; Kuang et al., 2017), allowing reliable waveform modeling. At regional and teleseismic distances, waveform inversion typically focuses on long-period body waves, surface waves, or W-phase signals, which are less sensitive to small-scale velocity heterogeneity (Dziewonski et al., 1981; Zhao & Helmberger, 1994; Zhu & Helmberger, 1996; Pasyanos et al., 1996; Tan & Helmberger, 2007; Kanamori and Rivera, 2008; Cesca et al., 2010). Beyond their reliance on accurate velocity models, waveform modeling and inversion approaches are computationally intensive, although recent machine-learning-based methods offer the potential to reduce this burden (e.g., Yang et al., 2021; Kuang et al., 2021). As a result, these waveform-fitting methods are most commonly applied to moderate to large earthquakes (e.g., M > 3).

An alternative and widely used class of methods avoids explicit waveform modeling, which relies on P- and S-wave first-motion polarities and/or S/P amplitude ratios (e.g., Reasenberg & Oppenheimer, 1986; Hardebeck and Shearer, 2002, 2003; Skoumal et al., 2024). Because amplitude ratios eliminate the need for absolute amplitude modeling (Kisslinger, 1980; Kisslinger et al., 1981), they reduce sensitivity to certain path effects. Using systematic grid-search approaches that incorporate uncertainty through parameter perturbations (e.g., the HASH algorithm; Hardebeck & Shearer, 2003), focal mechanisms can be statistically constrained with different levels of confidences. More recently, this traditional inversion process has been replaced by neural-network-based approaches (Song et al., 2025), resulting in more reliable solutions. These methods are particularly suitable for small to moderate earthquakes due to their lower computational demand and independence on waveform modeling. Nevertheless, several challenges remain: First, identifying first-motion polarities can be difficult, especially for low SNR (Signal-to-Noise Ratio) events or emergent arrivals. Although machine-learning-based picking methods have recently improved polarity identification (Ross et al., 2018; Uchide, 2022; Zhao et al., 2023), uncertainty remains significant for small events. In practice, polarity analysis primarily uses P-wave arrivals, while S-wave polarities are less commonly applied because of their generally lower quality. Second, amplitude ratios are affected by inaccuracies in velocity models and site effects, and therefore typically require empirical station corrections (Hardebeck & Shearer, 2003). Discrepancies between theoretical and observed P/S amplitude ratios often exceed a factor of two (Hardebeck & Shearer, 2003; Trugman, 2025), introducing uncertainty in inversion results.

Recent advances have demonstrated that both polarities and amplitude ratios can be jointly inverted for clusters of earthquakes using cross-correlation and matrix inversion techniques (e.g., Shelly et al., 2016; Shelly et al., 2022). By providing a larger number of more accurate inputs to HASH, these approaches significantly increase the number of solvable events and improve solution stability. However, they still reply on the HASH framework and inherit some limitations, such as the need for station corrections to achieve accurate solutions. More recently, Cheng et al. (2023a) proposed a new iteration of HASH that incorporates inter-event P/P and S/S amplitude ratios to refine the initial focal mechanism solutions obtained using HASH.

Similar to relative earthquake location methods (Waldhauser & Ellsworth, 2000), relative focal mechanism inversion approaches have been developed but have not been widely adopted due to various limitations. Dahm (1996) first developed a relative moment tensor inversion method that jointly solves a group of nearby earthquakes, thereby reducing dependence on Green's functions. This approach was later extended by Plourde & Bostock (2019), who introduced a generalized framework that incorporates S-waves and applies principal component analysis to measure precise relative amplitudes. More recently, Drolet et al. (2023) further improved the method by incorporating constraints from P-wave first-motion polarities. These ray theory-based relative inversion approaches solve focal mechanisms simultaneously for clusters of earthquakes, typically involving relatively complex computation procedures. In a different framework that does not rely on ray theory, relative focal mechanisms have also been investigated in waveform fitting based methods. Jia et al. (2022) adopted results from waveform inversion as prior information and applied Bayesian inversion to the body- and surface-wave amplitude ratios of an event pair, thereby refining the moment tensors of both events. Kuang et al. (2023) proposed using a well-constrained reference event to invert relative waveform differences between synthetic and observed data. Similar to other waveform-fitting-based approaches, both methods primarily target relatively large earthquakes (e.g., M > 3).

In this study, I propose a relative focal mechanism inversion method, named FocMecDR. Analogous to double-difference–based master-event relocation approaches (Wen, 2006; Wen & Long, 2009; Zhang & Wen, 2013), a well-constrained earthquake is selected as the reference event, and the focal mechanisms of nearby earthquakes within the cluster are solved relative to this reference. The relative polarities of P- and S-waves between the target and reference events, together with their S/P amplitude double ratios, are simultaneously used to constrain the double-couple focal mechanism solutions. In the following sections, I introduce the methodological framework, validate the approach using comprehensive synthetic tests and field data, and apply it to the foreshock sequence of the 2019 Ridgecrest earthquake.

## 2. Method

FocMecDR fully automatically constrains the focal mechanism of a target earthquake using a nearby reference event with a known focal mechanism. For small earthquakes, which can be approximated as point sources, source-time effects can largely be minimized by appropriate filtering within a particular frequency range. Because the target and reference events occur in close proximity, they share similar wave paths and Green's functions. This fundamental assumption is widely adopted and has been validated by related approaches, such as double difference for seismic location (e.g., Waldhauser & Ellsworth, 2000; Lin & Shearer, 2006) and template matching for seismic detection (e.g., Gibbons & Ringdal 2006; Zhang & Wen, 2015). Two types of information are incorporated in FocMecDR: relative polarities and S/P amplitude double ratios. FocMecDR reads segmented waveforms and automatically computes both constraints in memory without requiring additional input. I perform a grid search over strike, dip, and rake for the target event, minimizing a joint objective function defined by relative-polarity and S/P amplitude double-ratio misfits between the target and reference events (Equation 1 and Figure 1). The strike, dip, and rake are searched over the full solution space (strike: 0–360°, dip: 0–90°, rake: −180–180°) using a specified grid interval (e.g., 1° or 2°). The main objective function is defined as below:

$$\text{OBJ} = w1 * Mean(RCC_{p,sh,sv}) + w2 * Mean(DR_{psh}) + w3 * Mean(DR_{psv}) \qquad (1)$$

where w1, w2, and w3 are weighting coefficients with default values of 1.0. RCC represents the consistency measure between the theoretical and observed relative polarities for the P, SH, and SV phases between the event pair (Figure 1a). DR denotes logarithmic double-ratio misfits of SH/P and SV/P amplitudes between the event pair (Figure 1b).

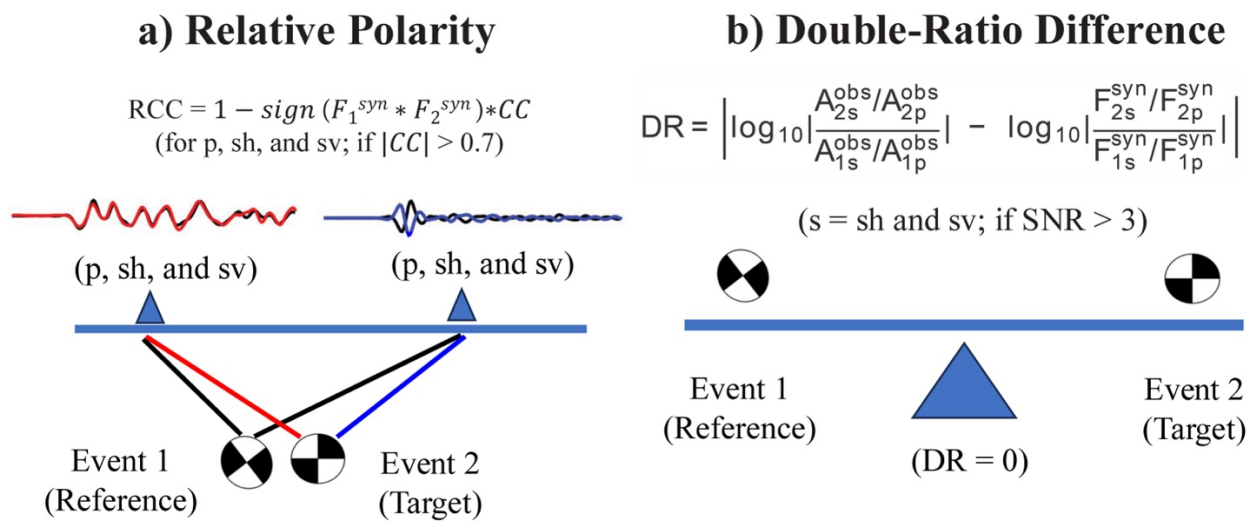

**Figure 1.** Schematic illustration of the objective function used in FocMecDR, including (a) consistency measure in the relative polarities of the P, SH, and SV phases, and (b) logarithmic double-ratio misfits of SH/P and SV/P amplitudes. Panel (a) illustrates the relative polarities between a pair of earthquakes with different focal mechanisms recorded at two common stations. Panel (b) demonstrates that only the optimal focal mechanism solutions can 'balance' the theoretical and observed double ratios of SH/P and SV/P amplitudes between the event pair. CC denotes the normalized cross-correlation coefficient for a seismic phase between a pair of events recorded at the same station. $F_1^{syn}$ and $F_2^{syn}$ represent the theoretical radiation-pattern coefficients of the seismic phase for the reference and target events (Aki & Richards, 2002), respectively, based on their focal mechanism solutions. $A_1^{obs}$ and $A_2^{obs}$ represent the observed amplitudes of the seismic phase for the reference and target events. For relative polarity estimation, a cross-correlation (CC) coefficient cutoff between event pairs (e.g., CC > 0.7) is adopted to reduce the risk of cycle skipping. For double-ratio calculations, a signal-to-noise ratio (SNR) cutoff (e.g., SNR > 3) is applied to ensure reliable amplitude measurements.

The relative polarities are automatically estimated for P, SH, and SV phases from cross-correlation (CC). Because the CC technique utilizes several seconds of waveform data rather than relying solely on the first-motion polarity of the initial arrival, it provides more robust polarity estimates and increases the number of usable measurements (Shelly et al., 2016; Shelly et al., 2022; Skoumal et al., 2023; Cheng et al., 2023a). In this framework, FocMecDR accesses the consistency or discrepancy of corresponding phases between an event pair without requiring the absolute polarity of each phase (Figure 1a). Ideally, when the CC between two co-located events approaches 1 and their relative polarities are consistent with theoretical predictions, the relative polarity misfit approaches zero (Figure 1a). Conversely, the relative-polarity misfit approaches 2 when their relative polarities are opposite to the theoretical predictions. In practice, the CC value may be lower due to slight spatial separation between the events, and this value can be used to weight the confidence assigned to each phase component (Figure 1a). A CC cutoff (e.g., 0.7) is recommended to reduce the risk of cycle skipping. Prior to polarity estimation, the horizontal components are rotated into radial (R) and transverse (T) components along the great-circle path. Waveforms are typically bandpass filtered within 1–8 Hz for regional studies to suppress noise and minimize source-dimension effects. This frequency band generally works well for datasets dominated by small to moderate earthquakes, although the optimal band may vary with event magnitude and source-receiver distance, which can consequently influence the appropriate CC cutoff. The P, SH, and SV phases are measured from the vertical (Z), T, and R components, respectively. The method also applies to stations equipped with only Z components, in which case only the P and SV phases recorded on the Z component are utilized (Kisslinger, 1980), with a smaller number of constrains. In addition, FocMecDR provides an option to incorporate manually picked high-confidence first-motion polarities of the P phase recorded on the Z component. If manual first-motion picks are provided, the misfit between the predicted and manually picked first motions is also incorporated into the total objective function. Because this option relies on absolute first-motion identifications, which are obtained using additional tools or through manual picking, it is not recommended for automated workflows but can be useful in particular situations when station coverage is limited or when few reliable relative polarities are available because of low waveform similarity.

FocMecDR adopts the double ratios of S/P amplitudes between the target event and a nearby reference event. Different focal mechanisms generate distinct radiation patterns at common seismic stations, which provide constraints on the double-couple solutions. For a pair of nearby earthquakes, the effects of velocity heterogeneity along the wave path and local site effects at the same station are largely canceled, and therefore no station corrections are required. Consistent with the relative polarity measurements, the absolute amplitudes of the P, SH, and SV phases are measured from the Z, T, and R components, respectively, for both the target and reference events, and are used to construct two amplitude ratios: SH/P and SV/P. Then I theoretically compute the far-field radiation-pattern coefficients for the P, SH, and SV phases (see Equations 4.89–4.91 in Aki & Richards, 2002) assuming a point-source dislocation, as the ratios of these coefficients are equivalent to the corresponding amplitude ratios. The source–receiver azimuths and takeoff angles required for computing the radiation-pattern coefficients are determined from the known hypocenters and a velocity model. For the optimal double-couple solution (i.e., strike, dip, and rake), the synthetic S/P double ratios derived from the radiation-pattern coefficients should match the observed S/P amplitude double ratios while also satisfying the relative polarity constraints between the event pair. Here, the P, SH, and SV phases independently form two double ratios, SH/P and SV/P. For stations with only Z components, the P and SV phases are measured from the Z component and the SH phase is not adopted.

Relative polarities and amplitude double ratios play complementary roles in FocMecDR. The double ratios are highly sensitive to the accuracy of the focal mechanism: even small changes in focal mechanism solutions can produce large variations in the S/P amplitude ratios at specific stations. In contrast, the relative polarities may remain unchanged if there are small changes in focal mechanism solutions when station density and azimuthal coverage are limited. Thus, amplitude double ratios are less sensitive to station azimuthal distribution, whereas relative (used in FocMecDR) polarities depend more strongly on azimuthal coverage, similar to absolute (e.g., used in HASH) polarities. On the other hand, theoretically, amplitude double ratios depend on the absolute amplitudes of seismic phases (Figure 1b) and therefore cannot independently distinguish between compressional and dilatational radiation patterns. This can lead to ambiguity between different faulting types, such as normal and thrust faulting or left-lateral and right-lateral strike-slip events. In practice, these differences might be partially reflected by waveform similarity (e.g., the CC value) and rupture directivity effects on amplitude ratios. To ensure reliable solutions, either prior constraints on the focal-mechanism search range, high-quality relative polarity measurements, or high-confidence manual absolute first-motion polarities are therefore necessary to uniquely determine the focal mechanism with correct faulting type. Among these strategies, sufficiently abundant, high-quality relative-polarity measurements enable the inversion process to be carried out fully automatically, without prior information or manual input.

FocMecDR operates on pairs of events but does not require selecting and fixing a particular reference event. Instead of choosing a single reference event, all eligible events within a specified region are used as potential references to solve the focal mechanism of the target event, and the solution associated with the reference event that yields the smallest misfit is retained. This procedure is similar to the strategy used in Match-and-Locate (e.g., Zhang & Wen, 2015). To evaluate misfits from different references with different station coverages, additional physical constraints are introduced to further assess the reliability of the solutions beyond relative polarities and amplitude double ratios. These include minimizing the station azimuthal gap for relative polarity measurements, minimizing the percentage of false relative polarities, and maximizing the percentage of usable relative polarities. These quantities are normalized between 0 and 1, making them comparable to the misfit terms from relative polarities and amplitude double ratios. Equal weighting is assigned to each term unless specific adjustments are required.

## 3. Synthetic Tests

Comprehensive synthetic tests are utilized to evaluate the performance of the FocMecDR method. A half-space velocity model with $Vp$ =6 km/s and $Vs$=3.5 km/s is adopted for waveform modeling. The earthquake depth is fixed at 10 km, and 12 stations evenly sample the source–receiver azimuth starting at 5° with an interval of 30° (Figure 2). Three fundamental faulting mechanisms (Lay & Wallace, 1995) are tested: vertical strike-slip (strike = 0°, dip = 90°, rake = 0°), vertical dip-slip (strike = 0°, dip = 90°, rake = −90°), and 45°-dipping thrust fault (strike = 0°, dip = 45°, rake = 90°). Waveform modeling is performed using the F-K program (Zhu & Rivera, 2002).

As a primary demonstration, the analysis focuses on a pair of co-located earthquakes with focal mechanisms corresponding to a 45°-dipping thrust event and a vertical dip-slip event. For the P phase, a window of 0.5 s before and 1.5 s after the P arrival is used. For the SH and SV phases, a window of 0.5 s before and 2.0 s after the arrival is used. These phase window settings are consistent with the later field tests, but they may vary depending on epicentral distance. A full grid search is performed over strike (0–360°), dip (0–90°), and rake (−180–180°), with grid intervals of 2° for strike and rake and 1° for dip, which are consistent with the later field verifications as well. Using the 45°-dipping thrust event as the reference, the vertical dip-slip objective solution is accurately recovered (Figure 2), with a computation time of less than 10 seconds on a MacBook Pro equipped with an Apple M3 Max chip. The data pre-processing and figure plotting take another tens of seconds. The relative polarities are fully consistent between the observations and predictions, showing expected positive and negative relative polarities at different phases and stations because of the focal mechanism difference (Figure 2a), which are more clearly illustrated by direct waveform comparisons (Figure 2c). The logarithmic double-ratio differences between synthetics and obversions approach zero, with negligible errors likely related to the accuracy of waveform modeling (Figure 2b). The total misfit shows robust convergence in strike, dip, and rake (Figure 2d). Multiple local minima arise from parameter equivalence, such as strike values of 0° and 360°, as well as the other mathematically indistinguishable auxiliary fault plane. With the same settings, FocMecDR performs consistently well when only vertical components are used, where the P and SV phases are extracted from Z components (Figure S1).

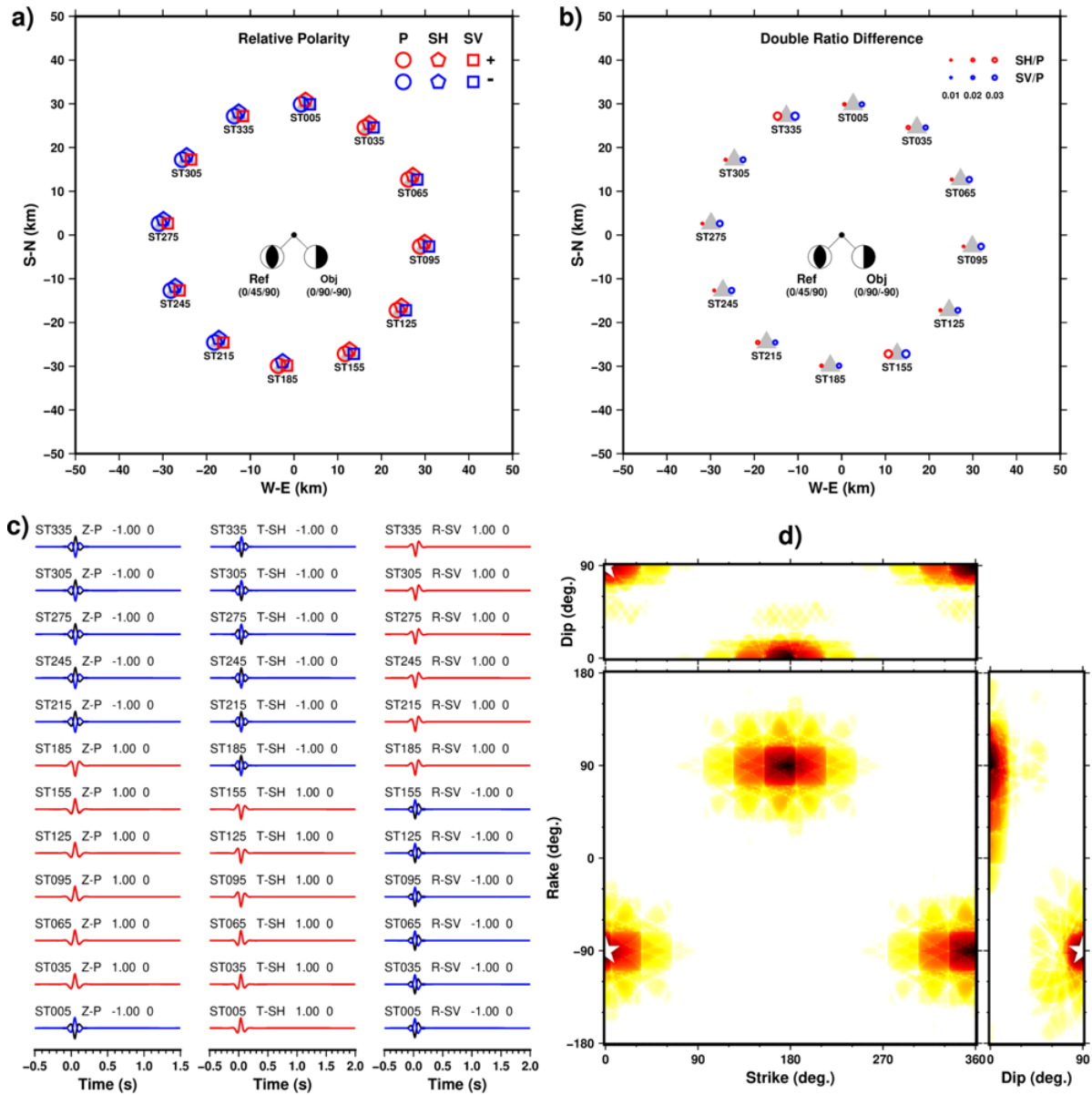

**Figure 2.** Synthetic test demonstrating that FocMecDR recovers the focal mechanism of a target event with vertical dip-slip using a co-located 45°-dipping thrust event as the reference. (a) Relative polarity consistency for different phases (P, SH, and SV) at different stations. Red indicates the expected consistent polarities, blue indicates the expected opposite polarities, and purple indicates unexpected polarity mismatches (not present in this case). (b) Logarithmic double-ratio differences for SH/P and SV/P, with circle size representing the misfit at each station. (c) Waveform comparisons for P, SH, and SV phases recorded on the Z, T, and R components. Black traces denote waveforms of the reference event, while red and blue traces represent waveforms of the target event with polarities consistent with or opposite to those of the reference event, respectively. The texts above each trace represent the station name, component-phase name, the maximum normalized CC within the window, and the corresponding phase shift associated with that maximum normalized CC. (d) Misfit distribution in the strike–dip–rake parameter space. The white star indicates the optimal focal mechanism corresponding to the minimum misfit.

To comprehensively evaluate different types of focal mechanisms, I designed two approaches to assess focal-mechanism uncertainties, focusing on the focal mechanism recovery of a target event using a 45°-dipping thrust event as the reference. The tests are separately conducted using both three-component and single-component recordings. In this analysis, the target event is not restricted to a specific focal mechanism. Instead, the focal mechanisms span all combinations of strike, dip, and rake, with strike ranging from 0° to 360° at an interval of 30°, dip from 0° to 90° at an interval of 15°, and rake from −180° to 180° at an interval of 30°. This configuration results in 864 independent focal-mechanism solutions. In the first approach, the reference event is assumed to have an accurate focal-mechanism solution (i.e., 45°-dipping thrust with strike = 0°, dip = 45°, and rake = 90°). The focal mechanism of each target case is then inverted and compared with the input focal mechanism solution using the Kagan angle (Kagan, 2007). The uncertainties of the target cases, 80% of the Kagan angles fall within 2.25°, likely limited by the grid-search resolution, with a few outliers reaching up to 7° (Figure 3a). In the second approach, to account for uncertainty in the focal mechanism of the reference event, the strike, dip, and rake of the reference are randomly perturbed by ±10° around their true values, resulting in an average Kagan angle perturbation of about 10°, but up to 20° (pink in Figure 3b). For each target focal-mechanism recovery, one randomly perturbed reference mechanism is used, and the Kagan angles between the predicted and input mechanisms are evaluated for all cases. The uncertainties of the target cases show that 80% of Kagan angles fall within 6.36°, and are mostly within 15°. Interestingly, the overall uncertainties of the recovered target events are smaller than those of the input reference solutions (Figure 3b), suggesting that FocMecDR does not fully inherit the uncertainties of the reference focal mechanisms. FocMecDR tends to guide the inversion toward the correct solutions for the target events, when the reference event introduces errors but still close the correct ones. Similarly, these two approaches were applied to the singe-component recordings (with Z only). The uncertainties show comparable distributions, but with slightly greater variation, as 80% of the Kagan angles fall within about 2° and 9° for the cases with fixed and perturbed reference focal mechanisms, respectively (Figure S2).

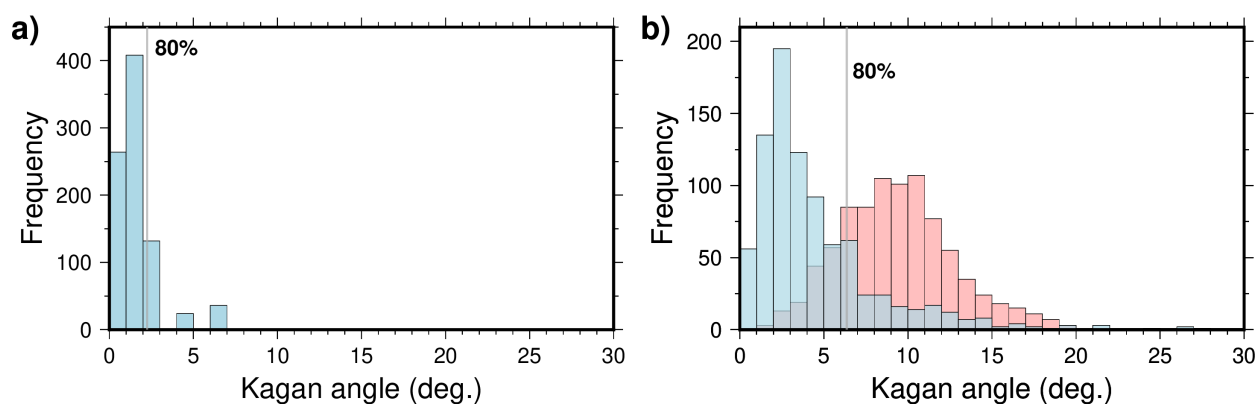

**Figure 3.** Uncertainty analysis using Kagan angles. (a) An accurate reference focal mechanism (45°-dipping thrust) is used to recover 864 independent different focal mechanism solutions, and the predicted mechanisms are compared with the input mechanisms using the Kagan angle (blue). (b) Same as in (a), but the reference focal mechanism is randomly perturbed within a Kagan angle of 20° (pink). The gray lines mark the values at which 80% of the Kagan angles are smaller.

Subsequently, I removed six stations and tested cases with a station azimuthal gap as large as 210° for both three-component (Figure S3) and vertical-component (Figure S4) configurations. The focal mechanisms are still reliably recovered, although the misfit convergence is not as smooth as with full azimuthal coverage because of less constrains (Figures 2 and S1). In addition, I tested the vertical strike-slip solution using the 45°-dipping thrust event as the reference with the same station configuration as Figure 2 (Figure S5). The focal mechanism is accurately recovered (Figure S5), although more local minima appear to accommodate auxiliary fault-plane solutions, as expected, because a vertical fault plane can exchange the footwall and hanging wall.

## 4. Field Verification

I adopted two cases to evaluate FocMecDR using field data from the 2019 Ridgecrest earthquake sequence. In the first case, I focused on a pair of nearby earthquakes with known focal mechanisms. One event was used as the reference to recover the focal mechanism of the other, and the results were then carefully analyzed. In the second case, I used an aftershock catalog of reported high-quality focal mechanisms and separated them into two groups. One group was used as the reference set to recover the focal mechanisms of the events in the other group. The consistencies and discrepancies between the FocMecDR results and the cataloged HASH focal-mechanism solutions are then analyzed and discussed.

### 4.1 Verification using a pair of antisimilar earthquakes during the 2019 Ridgecrest earthquake

Trugman et al. (2020) first reported antisimilar earthquake pairs during the 2019 Ridgecrest earthquake sequence and demonstrated waveform reversals at several stations. Such phenomena were later reported and further studied in this region and elsewhere (e.g., Wang and Zhan, 2020; Shearer et al., 2024; Cesca et al., 2024; Lu et al., 2025). In this case study, I focus on the event pair (Event ID: 38670527, M 2.01, 2019/07/30 with strike = 291°, dip = 84°, and rake = −154°; Event ID: 38674359, M 1.65, 2019/07/31 with strike = 188°, dip = 78°, and rake = −157°) shown in Figure 5 of Trugman et al. (2020), whose focal mechanisms are determined by the HASH method (Hardebeck & Shearer, 2003). The M 2.01 event, with its larger magnitude and more reliable focal mechanism, is used as the reference to recover the focal mechanism of the M 1.65 event.

I used public seismic stations located within 50 km, which recorded both of the two events in high quality. Three-component recordings are utilized when available; otherwise, vertical-component stations are included. The seismic velocity model from Shelly (2020) is adopted to compute theoretical arrival times and takeoff angles. For the P phase, a time window of 0.5 s before and 1.5 s after the theoretical arrival is used for cross-correlation and for measuring the absolute maximum amplitude from the Z component. For the SH and SV phases, a window of 0.5 s before and 1.5 s after the theoretical arrivals is used from the T and R components, respectively. For stations with only a Z component, the SH phase is not used, while the SV phase is measured from the Z component. All waveforms are band-pass filtered between 1 and 8 Hz prior to cross-correlation for relative polarity estimation and amplitude measurements. For reliable relative polarity estimation, only phases with CC > 0.7 are used, whereas amplitudes are used as long as their SNR exceeds 3. The strike, dip, and rake are searched over the full parameter space, consistent with the synthetic tests, using grid intervals of 2° for strike and rake and 1° for dip.

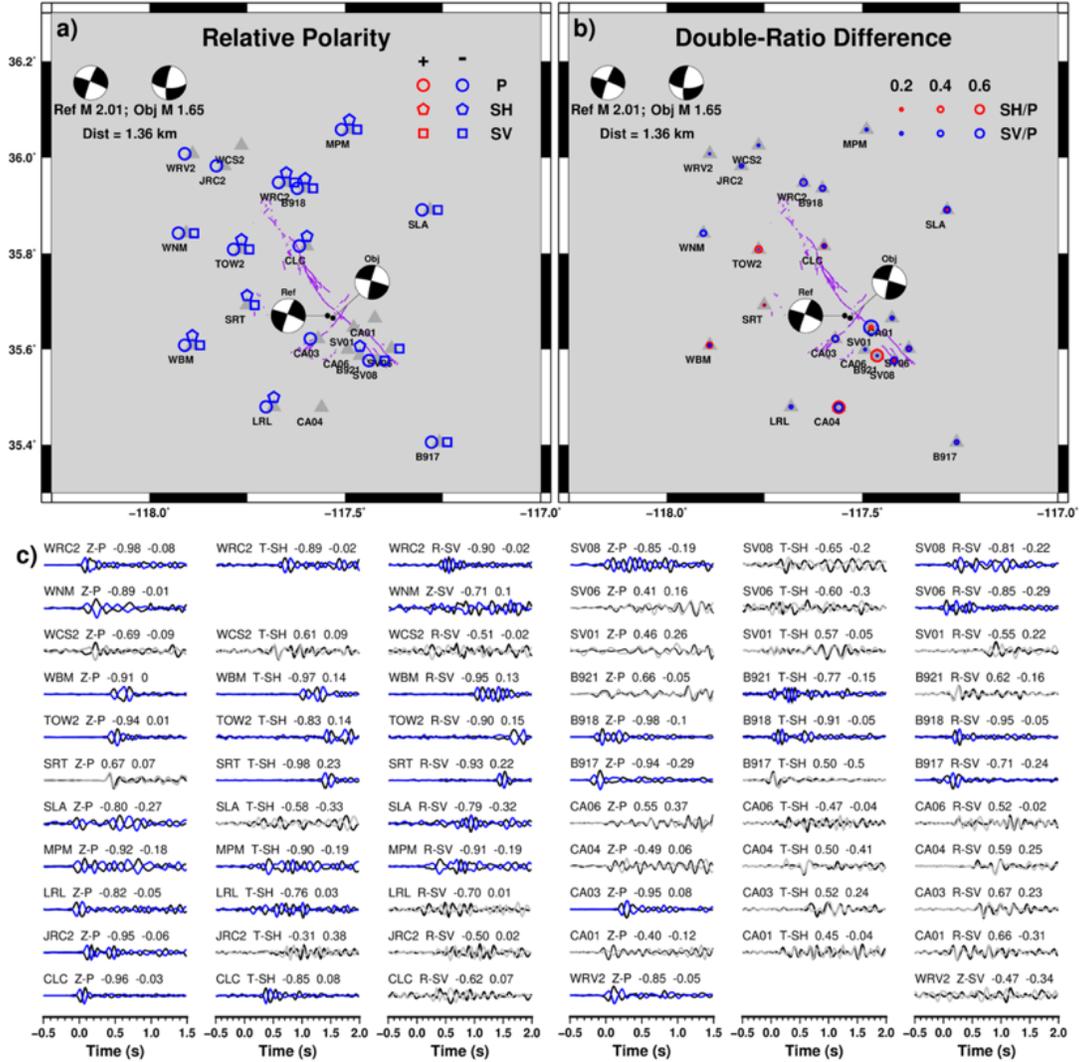

**Figure 4.** Field verification of FocMecDR using a pair of antisimilar nearby earthquakes from the 2019 Ridgecrest earthquake sequence reported by Trugman et al. (2020). (a) Relative polarities of P (circles), SH (pentagons), and SV (squares) for the event pair, with red indicating the expected consistent polarities (not present here), blue indicating the expected opposite polarities, and purple indicating unexpected polarity mismatches (not present here). (b) Logarithmic double-ratio misfits for SH/P (red circles) and SV/P (blue circles), with circle size proportional to the misfit at each station. (c) Waveform comparisons of the P, SH, and SV phases on the Z, T, and R components, respectively. Colored traces show the target event: red indicates phases expected to be consistent with the reference event (not present here), blue indicates phases expected to be opposite in polarity, and purple indicates unexpected polarity mismatches (not present here). The texts above each trace represent the station name, component-phase name, the maximum normalized CC within the window, and the corresponding phase shift associated with that maximum normalized CC. The focal mechanism solutions for the two events reported by Trugman et al. (2020) are shown in the upper left of panels (a) and (b). The recovered focal mechanism for the target event is shown together with its location, along with the reference event and its focal mechanism. In panels (a) and (b), mapped surface ruptures from Ponti et al. (2020) are shown in purple.

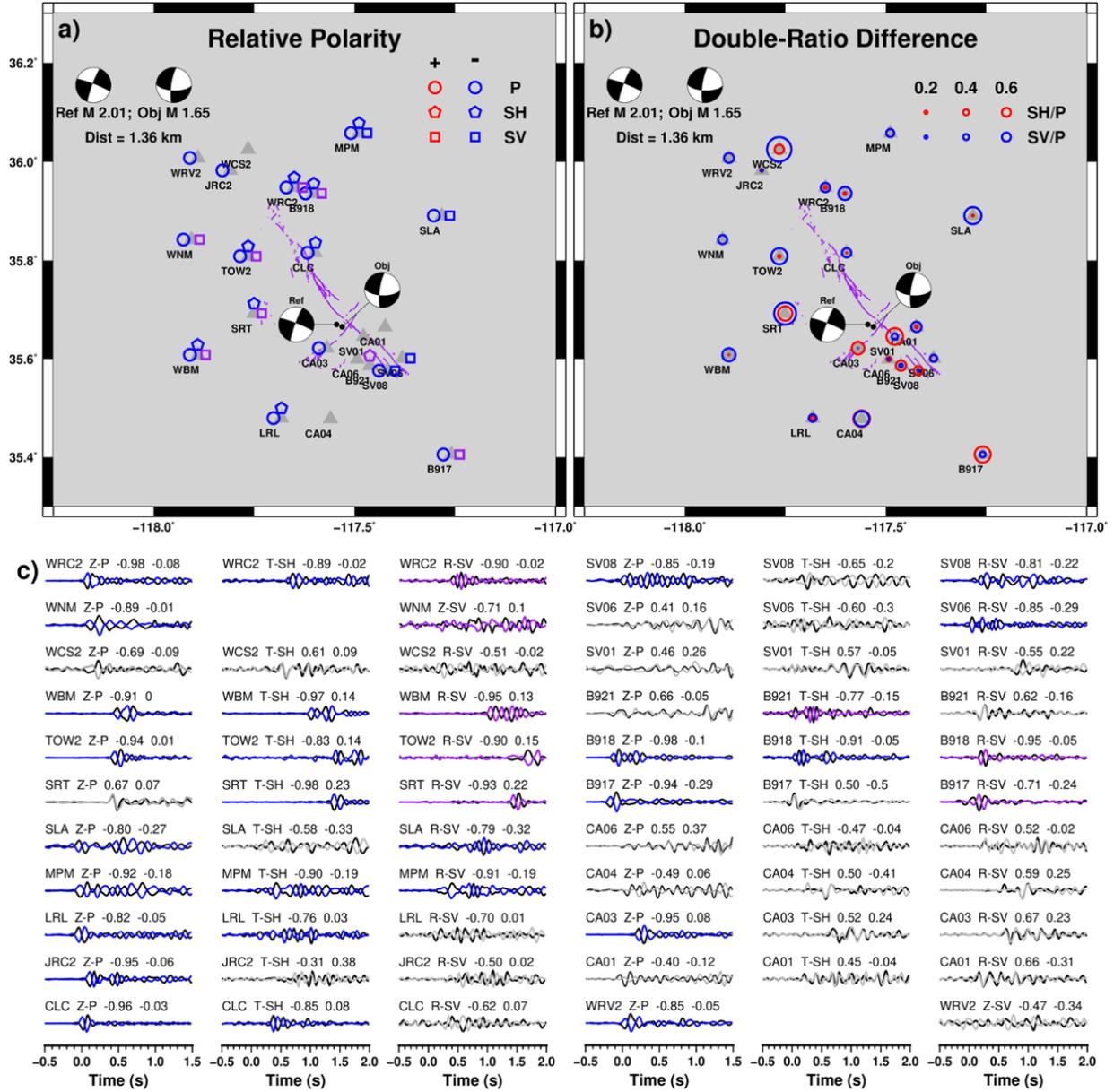

**Figure 5.** Similar to Figure 4, but using the HASH focal-mechanism solution from Trugman et al. (2020) for the target event. The purple symbols and purple traces in panels (a) and (c) denote phases with polarity mismatches. In panel (b) the logarithmic double-ratio misfits of SH/P and SV/P are significantly larger than those obtained when plotting with the FocMecDR solution.

FocMecDR recovers the focal mechanism of the M 1.65 event with strike = 194°, dip = 85°, and rake = 174°. All eligible relative polarities are negative, indicating an antisimilar rupture relationship between the two events (Figure 4a). The logarithmic double-ratio misfits of SH/P and SV/P are small, typically less than 0.4 (Figure 4b). Waveform comparisons of the P, SH, and SV phases between the two events are shown in Figure 4c, which provide additional details for visualizing the relative polarities. Components with phase CC < 0.7 are not used for relative polarity measurements. The reduced waveform similarity is likely related to the spatial separation of 1.36 km between the two events, which affects waveform similarity depending on the wave propagation path and azimuth, as well as the station locations relative to the nodal planes. The misfit distribution robustly converges to the optimal focal mechanism solution, with secondary minima corresponding to auxiliary fault planes (Figure S6).

The focal mechanism obtained with FocMecDR is consistent with the HASH solution (Figure 4), although discrepancies remain with a Kagan angle of 31°. It is therefore important to evaluate which solution better fits the observations. To do so, I recomputed the relative polarities and logarithmic double-ratio misfits using the HASH focal-mechanism solution to replace the FocMecDR solution (Figure 5). The consistency of the relative polarities and the magnitudes of the logarithmic double-ratio misfits provide direct evidence for assessing which solution better fits to the field data. Compared with the FocMecDR solution, the HASH solution introduces polarity mismatches and significantly larger double-ratio misfits (Figure 5). Notably, the P-wave relative polarities remain fully consistent, whereas mismatches occur for the SH and SV phases (Figure 5). This is likely because the HASH solution is primarily constrained by P-wave first-motion polarities, which can yield a first-order correct focal mechanism but may not provide an optimal fit to SH and SV observations. This case study not only demonstrates the effectiveness of FocMecDR, but also provides a practical approach for evaluating the correctness of focal-mechanism solutions.

## 4.2 Verification on the 2019 Ridgecrest earthquake

In this field verification, I used earthquakes associated with the 2019 Ridgecrest earthquake sequence from July to December 2019. Instead of focusing on the entire rupture fault zone, I selected the northwestern end of the rupture zone adjacent to the Coso geothermal field (Figure 6), which is characterized by active aftershocks and diverse faulting structures (Liu et al., 2020; Trugman et al., 2020; Wang and Zhan, 2020). The quality of focal mechanisms for both the reference and target events was cross-validated using A-quality HASH solutions from Southern California Seismic Network (SCSN; Yang et al., 2012) and Cheng et al. (2023b), who incorporated cataloged polarities and additional machine-learning-based polarities to improve both the accuracy and the number of solvable focal-mechanism solutions. Event locations and refined HASH focal mechanism solutions were taken from Chang et al. (2023). The dataset contains 147 earthquakes with magnitudes ranging from 0.8 to 4.9. Among these 147 events, 62 were selected as reference events, with both fault-plane uncertainties < 25°, number of first-motion polarities > 20 with misfit < 0.2, and number of S/P ratios > 20 with misfit ≤ 0.8 (see definitions in the HASH program). The remaining 85 events were treated as target events. The magnitudes of the reference dataset range from 1.7 to 4.5, whereas those of the target dataset range from 0.8 to 4.9. Only stations within 80 km that recorded both the reference and target events were used. All data-processing and parameter settings were as same as those described for the field verification in Section 4.1.

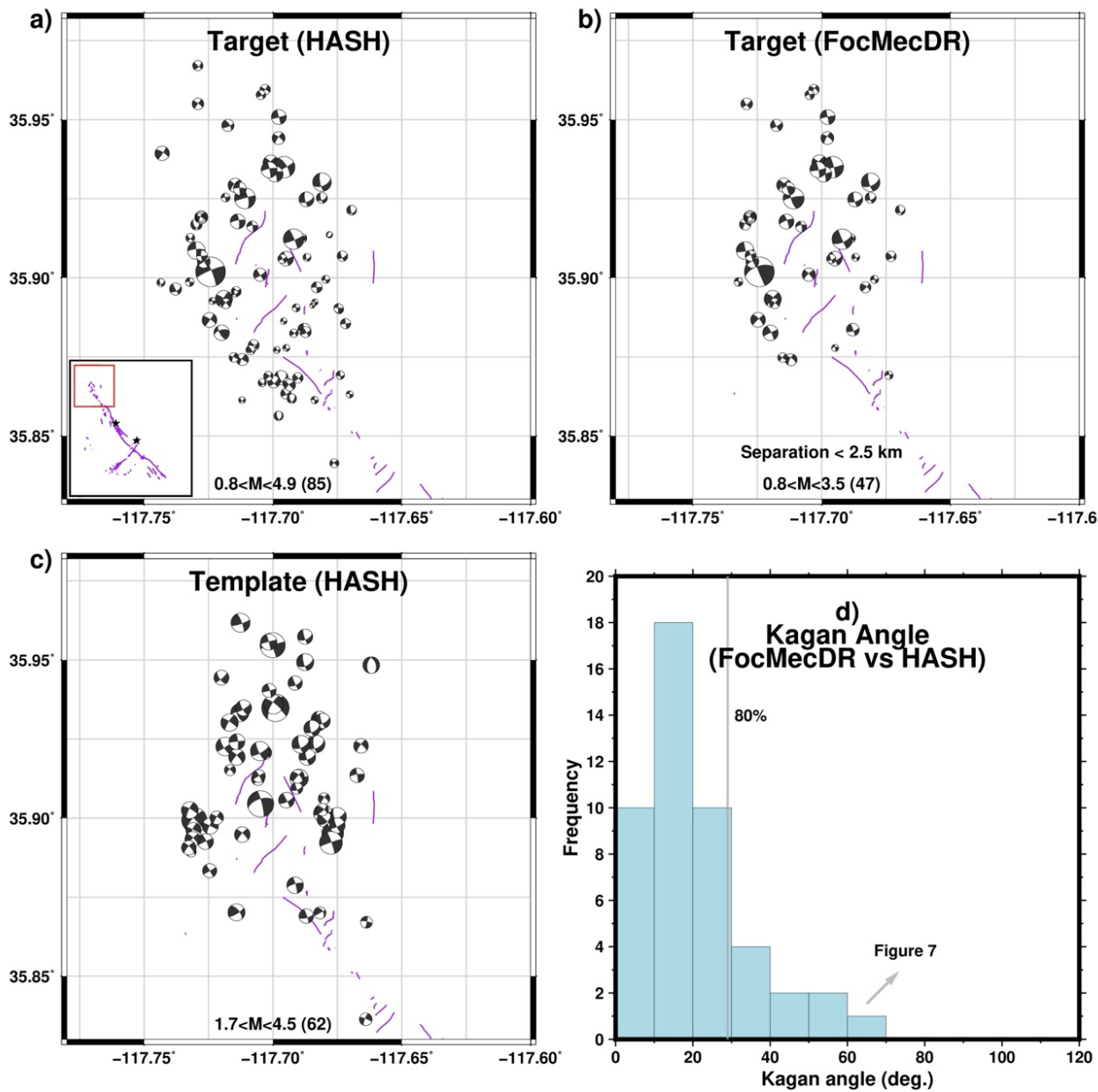

**Figure 6.** Field verification using a group of aftershocks at the northwestern end of the 2019 Ridgecrest earthquake rupture zone. (a) Spatial distribution of 85 target aftershocks with magnitudes ranging from 0.8 to 4.9, together with their HASH focal mechanism solutions. The mapped surface ruptures from Ponti et al. (2020) are shown in purple. The inset shows the broader rupture zone, with stars indicating the two mainshocks, Mw 6.4 and Mw 7.1. (b) Spatial distribution of the 47 resolved target events from panel (a) using FocMecDR. These events have nearby reference events within 2.5 km and magnitudes ranging from 0.8 to 3.5. (c) The reference dataset consists of 62 foreshocks with high-quality A-class HASH focal mechanism solutions and magnitudes ranging from 1.7 to 4.5. (d) Kagan angles between the FocMecDR and HASH solutions for the 47 common target events in panel (b), with 80% of the events have Kagan angles less than 29°. The reference-target event pair with the largest Kagan angle (~70°) is discussed in Figure 7.

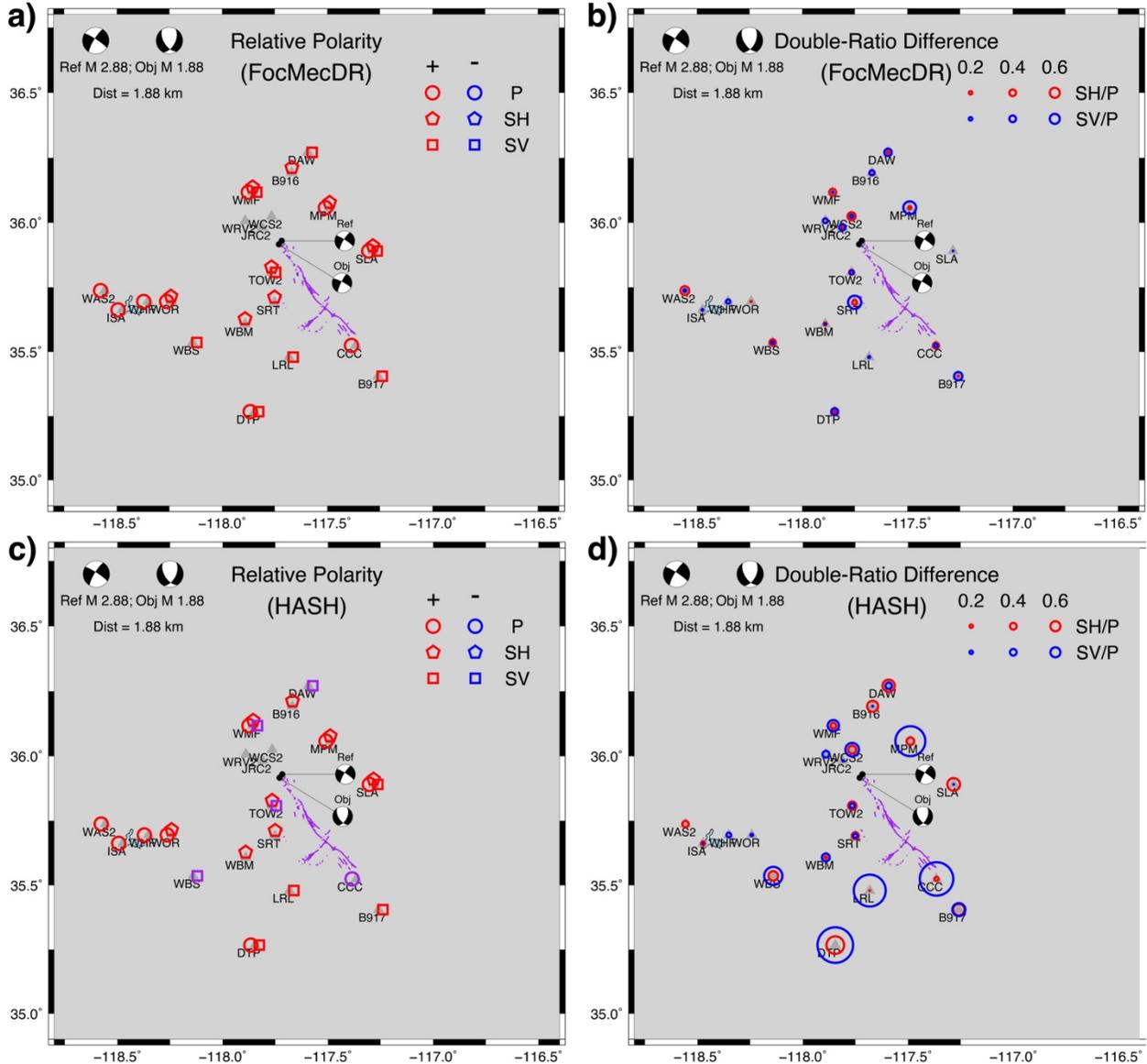

**Figure 7.** The relative polarities and double-ratio differences between the M 2.88 reference event and the M 1.88 target event. (a) Relative polarities of P (circles), SH (pentagons), and SV (squares) for the event pair, with red and blue indicating expected consistent and opposite relative polarities, respectively (purple would indicate mismatched relative polarities, although none are present here). (b) Logarithmic double-ratio misfits for SH/P (red circles) and SV/P (blue circles), with circle size proportional to the misfit at each station. In panels (a-b), the focal mechanism solution of the target event is from FocMecDR. (c–d) Similar to (a–b), except that the input solution is from HASH rather than FocMecDR. The purple symbols in panel (c), together with the enlarged SH/P and SV/P double-amplitude ratios, indicate the large errors produced by the HASH solution. In all panels, the HASH focal mechanisms are shown in the upper left. The used focal mechanism for the target event is shown together with its location, along with the reference event and its focal mechanism. In all panels, mapped surface ruptures from Ponti et al. (2020) are shown in purple.

Here, I did not select reference events specifically for each target event. Instead, for each target event, FocMecDR attempted to use the five nearest reference events and then retained the focal mechanism solution with the smallest misfit. This misfit considers contributions from relative polarities, S/P double ratios, and other physical factors such as station azimuthal coverage, the percentage of false relative polarities, the proportion of usable relative polarities (see Methods). This automatic selection strategy may also preferentially select event pairs with similar magnitudes and smaller separations, which optimizes the cross-correlation values and the proportion of usable phases. An empirical separation threshold of 2.5 km between reference–target event pairs is applied to ensure the validity of the empirical Green's function concept adopted in FocMecDR, resulting in focal-mechanism solutions for 47 of the 85 target earthquakes. The 38 unresolved events lack nearby reference events; they include scattered events to the south and northwest of the target region, as well as a group of events at 117.69°W, 35.86°N. The 47 FocMecDR solutions are broadly consistent with the refined HASH results of Chang et al. (2023) (Figures 6a and 6b), with 80% of the Kagan angles being less than 29° and the maximum reaching 70° (Figure 6d).

To verify the source of the mismatch between the solutions by FocMecDR and HASH, I focused on the reference–target event pair consisting of the reference event (Event ID: 38572695, 2019/07/15, M 2.88, strike = 303°, dip = 82°, rake = −168°) and the target event (Event ID: 38475063, 2019/07/07, M 1.88), which yields the largest Kagan angle of 70°. The two events are separated by 1.88 km and were recorded by 20 eligible stations within 80 km (Figure 7). The FocMecDR focal mechanism solution shows consistent relative polarities as the reference event among all eligible phase pairs (i.e., CC > 0.7; Figures 7a and S7), with the misfits of SH/P and SV/P double ratios are mostly within 0.4 (Figure 7b). However, when the focal mechanism solution is replaced by the HASH solution, five of the 25 eligible relative polarities become mismatched (Figures 7c and S8), including one for the P phase (station CCC) and four for the SV phase (stations DAW, TOW2, WBS, WHF, and WMF). In addition, the SH/P and SV/P double-ratio misfits increase significantly at most stations, particularly for the SV/P double ratios at four stations (CCC, DTP, LRL, MPM), reaching values of up to 2 (Figure 7d). There are 12 of 20 stations with reported first motions from SCSN (CCC, +1; ISA, +1; MPM, +1; SRT, -1; TOW2, -1; WAS2, -1; WBM, -1; WHF, -1; WOR, 1; WRV2, 1; B916, -1; B917, +1), two of which can be confidently identified as incorrect. Station CCC was reported as having a positive P-wave first motion, but it should be negative, whereas station WAS2 was reported as having a negative P-wave first motion, but it should be positive (Figure S8). The absence of S-wave polarities causes the HASH solution to fit the S-wave relative polarities less well, particularly for the SV phase (Figures 7c and S8). Although Cheng et al. (2023b) used a deep-learning first-motion classifier to add additional polarities for HASH inversion, they retained the original cataloged polarities, so those errors are also carried into their results.

Here, the underlying assumption is that the Kagan angle is primarily attributed to the target event, with only a relatively minor contribution from the reference event, because a more reliable focal mechanism solution is typically chosen for the reference event. Directly verifying the accuracy of the reference event is challenging. To address this, I designed an approach to indirectly evaluate the reliability of the reference event. Specifically, I treated the M 2.88 reference event as a target event and inverted its focal mechanism using the same procedure, with the remaining events from the reference dataset serving as references. An M 2.70 earthquake (Event ID: 38508743; 2019/07/09; strike = 317°, dip = 78°, rake = −166°), with a separation of 1.28 km, was retained as the reference event because it yielded the smallest misfit. This resulted in a focal mechanism solution (strike = 310°, dip = 71°, rake = 174°) for the M 2.88 "target" event (Figure S9), with a Kagan angle of 24° relative to its cataloged HASH focal mechanism (strike = 303°, dip = 82°, rake = −168°). This Kagan angle represents an upper bound on the uncertainty of the M 2.88 reference event, as it may also include uncertainty introduced by the M 2.70 reference event. Therefore, the cataloged HASH solution of the reference event is considered reliable, and the misfit primarily reflects errors associated with the target event.

## 5. Applications to the foreshocks of the 2019 M 6.4 Ridgecrest earthquake

About two hours before the 2019 Mw 6.4 Ridgecrest mainshock, foreshocks were widely reported (Shelly, 2020; Liu et al., 2020). Using the Match-and-Locate method, Liu et al. (2022) expanded the cataloged number of foreshocks from 9 to 40 and obtained high-resolution relative locations. Coulomb stress analysis indicates that the events after the largest M 4.0 foreshock and the Mw 6.4 mainshock can be explained by cascade triggering (Liu et al., 2022). In this application, I utilized the largest M 4.0 foreshock as the reference (Event ID: 38443095), which has well-constrained focal mechanism (SCSN; Quality A, strike=140°, dip=87°, rake=170°), to relatively determine the focal mechanisms of the rest foreshocks with magnitudes ranging from -0.23 to 2.27 (Liu et al., 2022).

Seismic stations within 50 km that recorded both the reference and target events were used in FocMecDR. To ensure reliability for small-magnitude events, FocMecDR only solves events for which at least five stations recorded high-quality phases for both events (i.e., SNR > 3), resulting in 15 of the 39 target events meeting this criterion (Figure 8). Less stringent criteria may still work, but they are likely to introduce large errors and would require manual evaluation based on the overall misfit and misfit convergence. A 1° interval for strike, dip, and rake was adopted for a finer-resolution search of the full parameter space. The data filtering (i.e., 1–8 Hz) and CC threshold (i.e., 0.7) were the same as those used in the previous field verifications.

All 15 events have their focal mechanisms successfully recovered, showing strike-slip faulting that is highly consistent with the M 4.0 reference event. Along with the template event, the 16 foreshocks exhibit a consistent strike-slip faulting pattern. By choosing the fault plane consistent with the M 4.0 reference event (i.e., strike = 140°), the mean strike, dip, and rake of the 16 foreshocks are 139.6°, 86.7°, and 170°, respectively, with standard deviations of 1.8°, 1.6°, and 5.4°. Their corresponding mean P-axis azimuth is 84.75° ± 1° (or equivalently 264.75° ± 1°) (Figure 8), indicating a highly uniform stress state during the mainshock nucleation process.

An example is shown for a reference–target event pair consisting of the M 4.0 reference event and an M 0.2 target event (Figure S10), which displays consistent relative polarities for the P, SH, and SV phases, with SH/P and SV/P double-ratio misfits of less than ~0.4, except for the SV/P ratio recorded at station SLA, potentially because its event–station azimuth is close to one of the focal nodal planes. Although the magnitude difference between the reference and target events spans about four orders of magnitude (M 4.0 vs. M 0.2), their waveforms still show similarity after band-pass filtering between 1 and 8 Hz, enabling the use of FocMecDR.

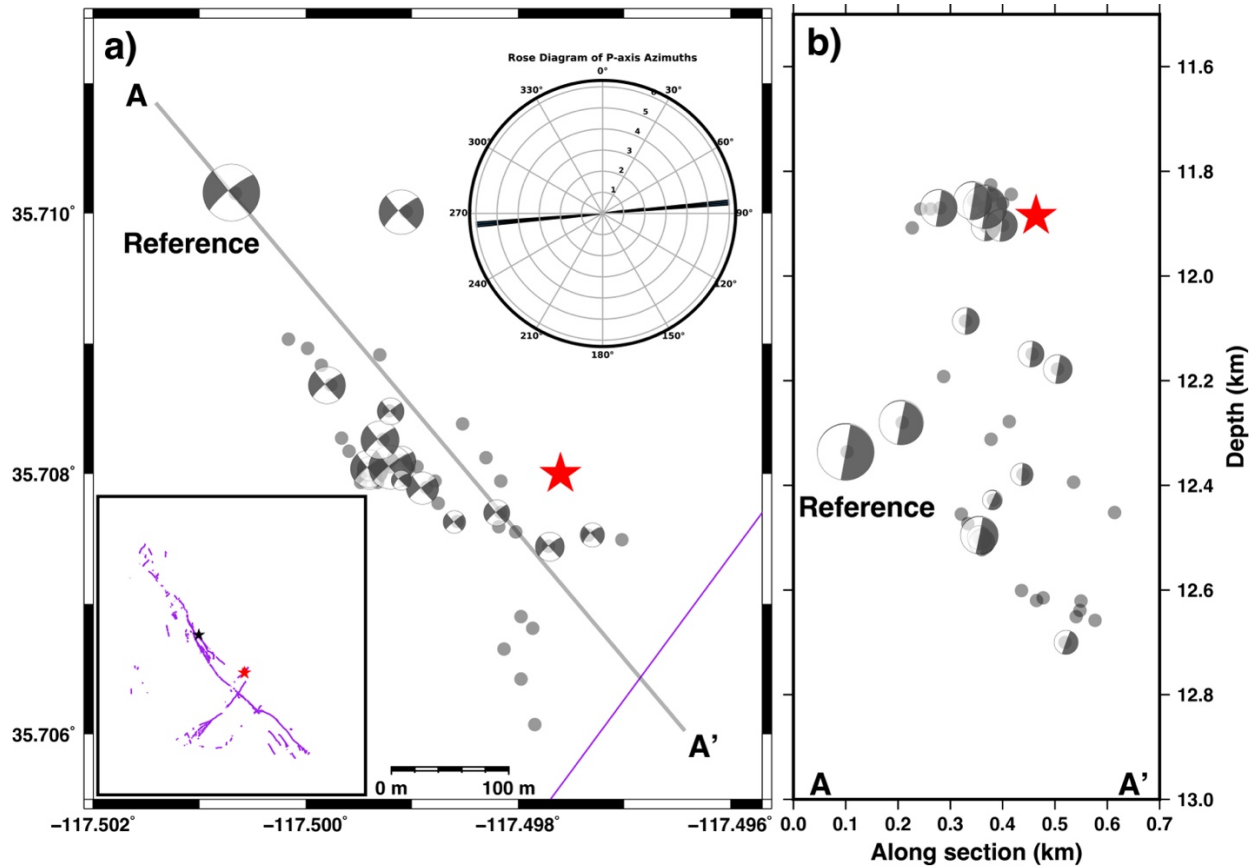

**Figure 8.** Application of FocMecDR to foreshocks preceding the 2019 Ridgecrest Mw 6.4 mainshock. (a) The largest foreshock (M 4.0) with a reliable HASH focal mechanism solution is selected as the reference event. Focal mechanisms for 15 of 39 events are resolved using FocMecDR; these events were recorded at more than five common stations by both the reference and target events. The inset shows the broader rupture zone, with red and black stars indicating the Mw 6.4 and Mw 7.1 mainshocks, respectively. Mapped surface ruptures from Ponti et al. (2020) are shown in purple. Gray dots denote all foreshocks from Liu et al. (2022), and the red star marks the Mw 6.4 mainshock. Beach balls represent the focal mechanisms of the M 4.0 reference event and the 15 resolvable foreshocks. The upper right panel shows a rose diagram of P-axis azimuths for the 15 resolvable foreshocks together with the reference event. (b) Depth projection along profile A–A′ in panel (a) for all 40 foreshocks, together with focal mechanisms of the 15 resolvable events and the reference event.

Here, we infer a localized and relatively uniform local stress field from the foreshock sequence during the approximately two hours preceding the 2019 Ridgecrest mainshock. The observed uniformity of focal mechanisms may reflect a critical preparatory stage before the mainshock. In contrast, fault zones that are not yet close to rupture may host earthquakes with more variable focal mechanisms, as is often observed in some post-mainshock sequences. I therefore hypothesize that, as stress accumulates, a fault zone may progressively evolve toward a more coherent local stress field and produce earthquakes with increasingly consistent focal mechanisms. This interpretation is broadly consistent with studies of earthquake nucleation and stress evolution: immediate foreshocks can show highly similar waveforms and common cascading rupture development (Meng & Fan, 2021), large earthquakes can release much of the spatially uniform component of the background shear stress (Hauksson, 1994), and stress rotations have been observed before and after major earthquakes (Yoshida et al., 2012; Hasegawa et al., 2012). At the same time, contrasting behaviors are also well documented (Peng & Lei, 2025). In some sequences, foreshocks and mainshocks have different focal mechanisms, such as the 2018 Hualien and 1975 Haicheng earthquakes (Jian et al., 2019; Jones et al., 1982), or rupture different fault branches, such as the 2010 Yushu and 2021 Yangbi earthquakes (Chuang et al., 2023; Liu et al., 2022). These differences may reflect the influence of fault geometry (Guo et al., 2023), fault orientation relative to the stress field (Liu-Zeng et al., 2024), frictional and strength properties (Tal et al., 2017), and structural complexity within the damage zone (Savage et al., 2017). Thus, the specific link between increasing focal-mechanism uniformity and rupture readiness remains to be tested, particularly regarding the conditions under which it holds. To fully test this idea, future work should examine additional foreshock sequences and compare them with long-term background seismicity and aftershock sequences.

## 6. Discussions

Instead of relying on absolute polarities and S/P amplitude ratios, FocMecDR uses relative polarities and S/P double ratios. Relative-polarity measurements are obtained fully automatically through cross-correlation techniques, providing high-quality polarity constraints for the P, SH, and SV phases while increasing the number of usable observations. The S/P double ratios introduced in this work, including SH/P and SV/P, are conceptually similar to the double-difference approach in relative earthquake location, as they largely cancel the effects of unknown velocity structure and site response, thereby improving measurement precision. These new constraints allow FocMecDR to determine high-precision double-couple focal mechanisms and, consequently, to better constrain regional stress field. In addition, such high-resolution relative focal-mechanism solutions enable nucleation processes to be investigated not only through temporal changes in precise foreshock locations (Ellsworth and Bulut, 2018), but also through more detailed temporal variations in the local stress field.

FocMecDR is a relative focal-mechanism inversion method and therefore requires well-constrained reference focal mechanisms. Synthetic tests show that uncertainties in the reference-event focal mechanisms do propagate into the target-event solutions, although less strongly than might be expected. For this reason, high-quality focal mechanisms (e.g., A-quality solutions from HASH) are recommended for the reference events. FocMecDR is not meant to replace conventional focal-mechanism inversion methods such as HASH; rather, it provides an alternative and independent tool for detailed analyses, particularly of temporal focal-mechanism evolution and for verifying the correctness of focal-mechanism solutions.

At present, FocMecDR uses constraints derived from individual reference–target event pairs. However, these constraints could be generalized to an earthquake cluster, analogous to the double-difference location method hypoDD (Waldhauser and Ellsworth, 2000). One important caveat is that the focal mechanisms within the cluster should not be identical, but instead should be sufficiently diverse to provide independent constraints (Dahm, 1996), in which case no reference events would be required. For such an earthquake cluster, FocMecDR could solve for focal mechanisms using more efficient inversion schemes, such as least-squares inversion and machine learning. This will be explored in future work.

## 7. Conclusions

I developed a new relative focal-mechanism inversion method, FocMecDR, which uses a well-constrained focal-mechanism solution as a reference to solve for nearby target events based on their relative polarities and double ratios of S/P amplitudes. FocMecDR utilizes the relative polarities of the P, SH, and SV phases, as well as double ratios of SH/P and SV/P amplitudes between pairs of events. Similar to the double-difference approach, FocMecDR largely cancels unknown effects from velocity structure and station sites, thereby achieving high-precision focal-mechanism inversion. Although the accuracy of FocMecDR depends on the reference events, synthetic tests suggest that this dependence is weaker than might be expected. FocMecDR offers a useful approach for tracking detailed temporal stress variations and provides a new strategy for evaluating the accuracy of focal-mechanism solutions between pairs of events. When applied to foreshocks of the 2019 M 6.4 Ridgecrest sequence, highly uniform focal mechanisms are observed, suggesting localized and homogeneous stress conditions that may indicate rupture readiness and critical preparatory processes prior to the mainshock, although this hypothesis requires validation through future work. FocMecDR therefore provides a complementary perspective for studying earthquake nucleation through detailed stress evolution.

## Data Availability Statement

All seismic waveform data used in this study were obtained from the Earthscope Inc. Data Management Center (DMC), formerly known as Incorporated Research Institutions for Seismology (IRIS) DMC. The SCSN HASH focal mechanisms and polarities used for method verification are from the Southern California Earthquake Data Center. https://doi.org/10.7909/C3WD3xH1. Maps presented in this study were generated using the Generic Mapping Tools (GMT; Wessel et al., 2013). The FocMecDR software will be released upon acceptance of the manuscript. Early access is available upon request via email.

## Acknowledgments

The author is grateful to Lianxing Wen, William Ellsworth, and Gregory Beroza for their inspiration and encouragement. The S/P double ratio concept was initiated after the publication of Zhang and Wen (2015). The primary coding was completed during the COVID-19 pandemic, and preliminary results (based on vertical component recordings) were presented at the 2023 Seismological Society of America Annual Meeting (Zhang, 2023). The author acknowledges the support received from the Natural Sciences and Engineering Research Council of Canada Discovery Grant (RGPIN-2019-04297).

# Supplementary Information

This document contains ten figures supporting additional synthetic tests, analysis, and discussions presented in the main manuscript.

Supplemental figures are presented as:

- Figure S1-S5: Additional synthetic tests and analysis
- Figure S6: Misfit distribution for the application of FocMecDR to a pair of antisimilar earthquakes
- Figure S7-S8: Waveform comparisons using the FocMecDR solution and the HASH solution for the event pair with the largest Kagan angle
- Figure S9: FocMecDR solving the M 2.88 event using a nearby M 2.7 reference event
- Figure S10: An example showing that FocMecDR solves an M 0.2 foreshock using the largest foreshock (M 4.0) as the reference

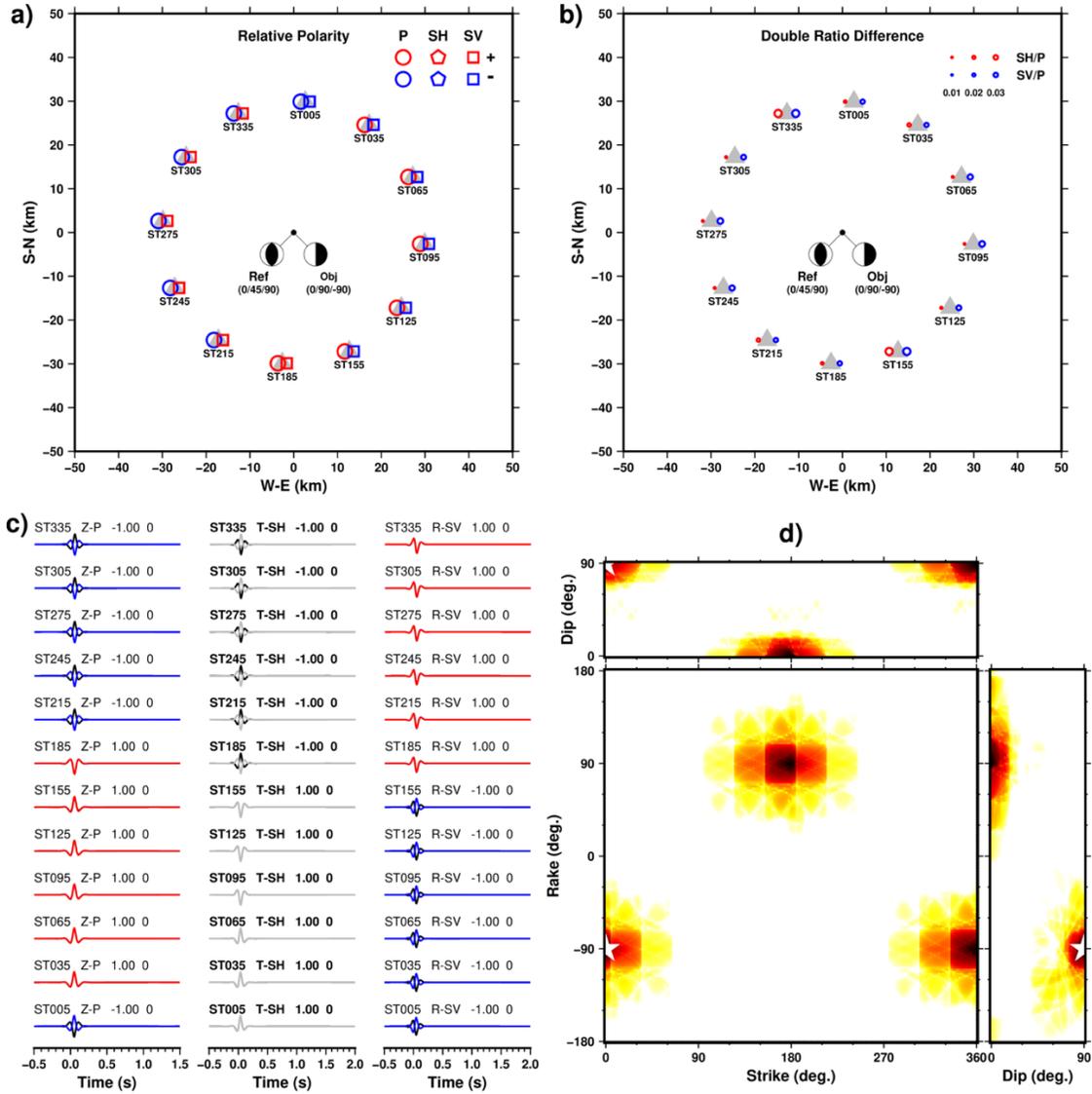

**Figure S1.** Synthetic test demonstrating that FocMecDR recovers the focal mechanism of a target event with vertical dip-slip using a co-located 45°-dipping thrust event as the reference. (a) Relative polarity consistency for different phases (P, SH, and SV) at different stations. Red indicates the expected consistent polarities, blue indicates the expected opposite polarities, and purple indicates unexpected polarity mismatches (not present in this case). (b) Logarithmic double-ratio differences for SH/P and SV/P, with circle size representing the misfit at each station. (c) Waveform comparisons for P, SH, and SV phases recorded on the Z, T, and R components. Black traces denote waveforms of the reference event, while red and blue traces represent waveforms of the target event with polarities consistent with or opposite to those of the reference event, respectively. The texts above each trace represent the station name, component-phase name, the maximum normalized CC within the window, and the corresponding phase shift associated with that maximum normalized CC. (d) Misfit distribution in the strike–dip–rake parameter space. The white star indicates the optimal focal mechanism corresponding to the minimum misfit. This figure is the same as Figure 2 in the main text, except that only vertical-component recordings are used here.

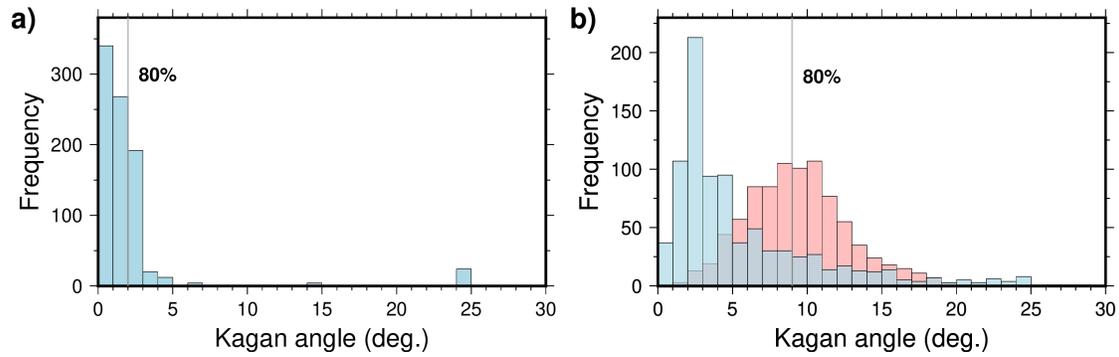

**Figure S2.** Uncertainty analysis using Kagan angles. (a) An accurate reference focal mechanism (45°-dipping thrust) is used to recover 864 independent different focal mechanism solutions, and the predicted mechanisms are compared with the input mechanisms using the Kagan angle (blue). (b) Same as in (a), but the reference focal mechanism is randomly perturbed within a Kagan angle of 20° (pink). The gray lines mark the values at which 80% of the Kagan angles are smaller. This figure is the same as Figure 3 in the main text, except that only vertical-component recordings are used here.

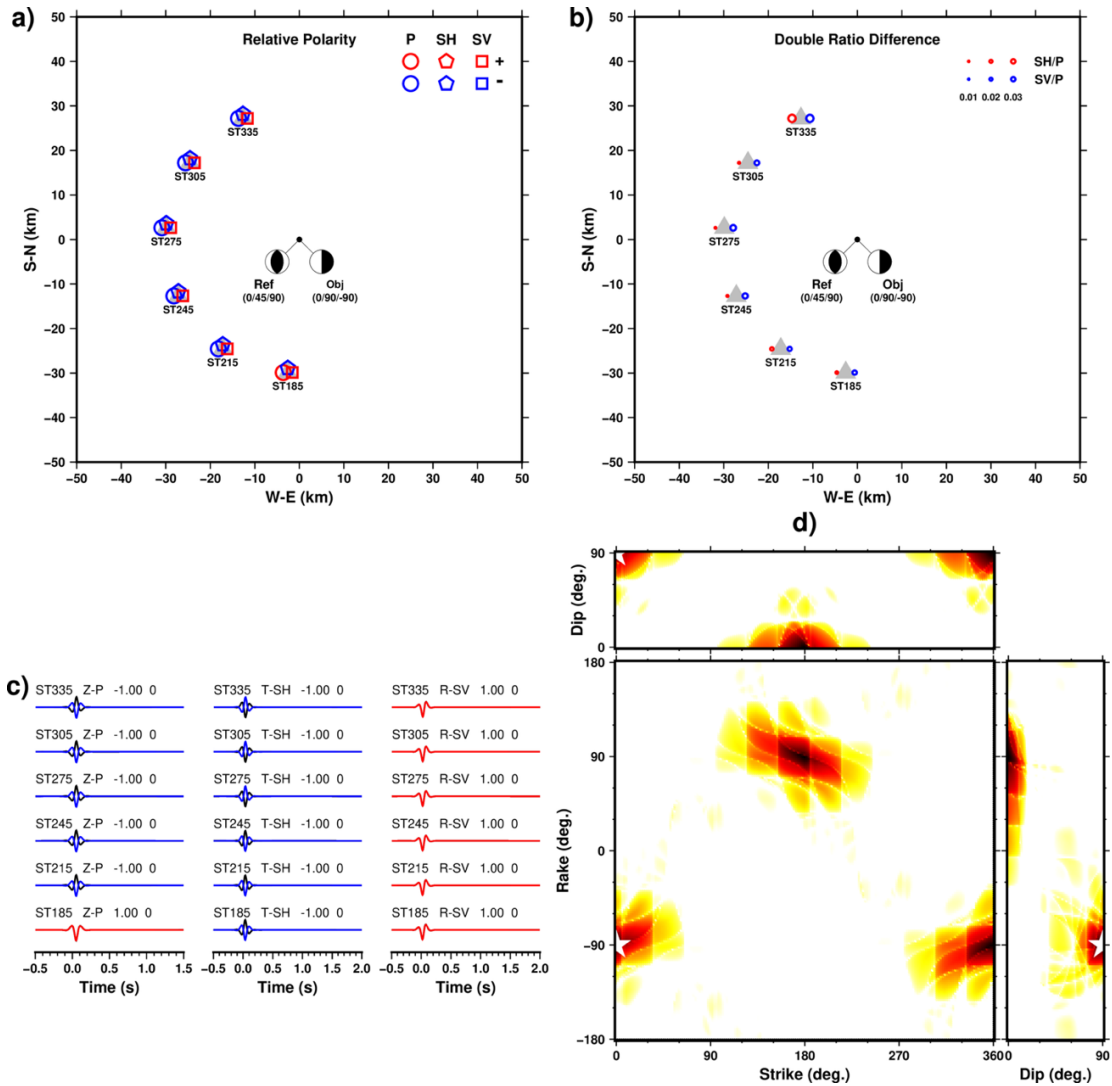

**Figure S3.** Similar to Figure 2 in the main text, except that 6 of 12 stations are removed, resulting in an azimuthal gap of 210°.

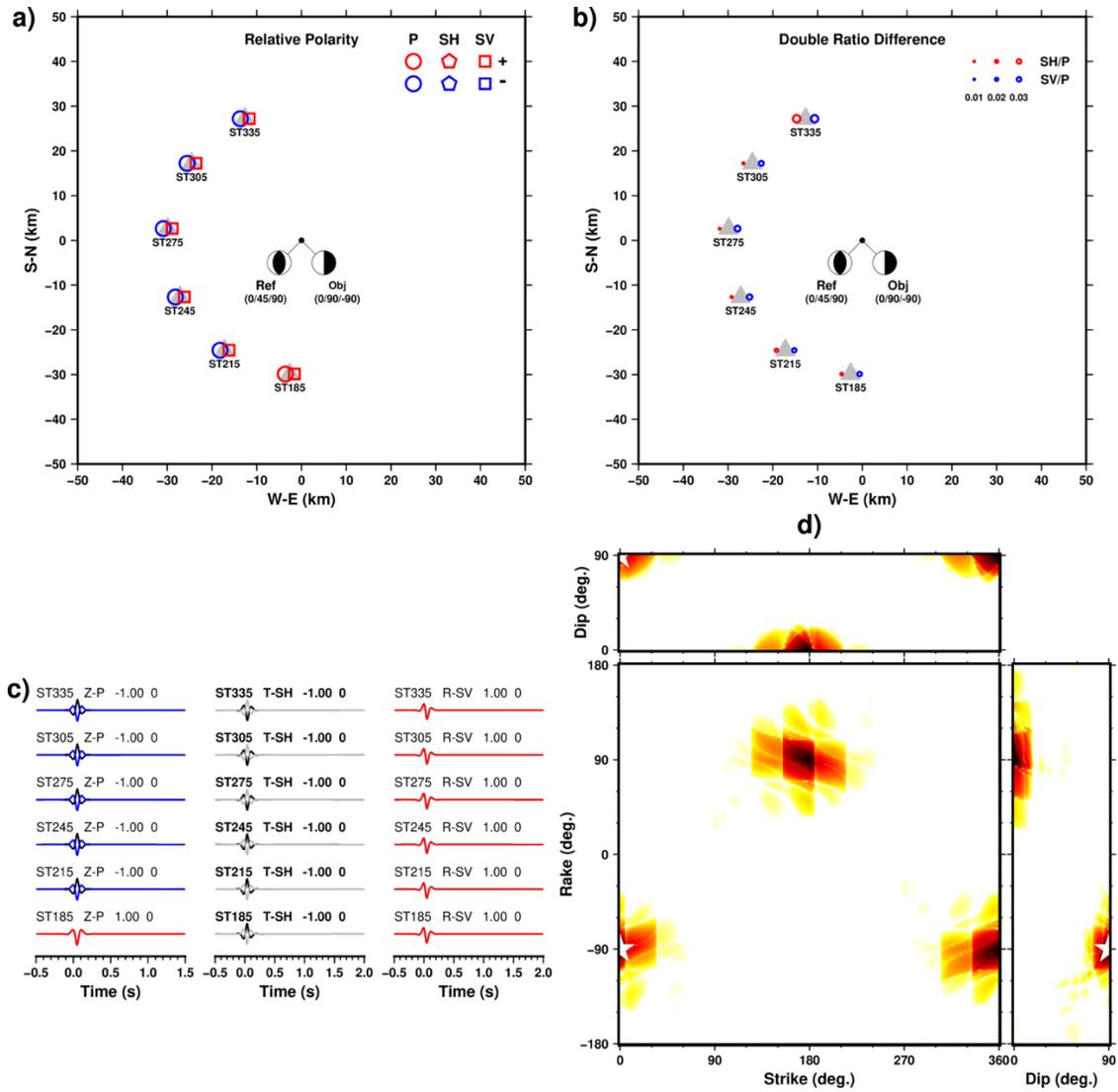

**Figure S4.** Similar to Figure 2 in the main text, except that 6 of the 12 stations are removed, resulting in an azimuthal gap of 210°, and only vertical-component recordings are used.

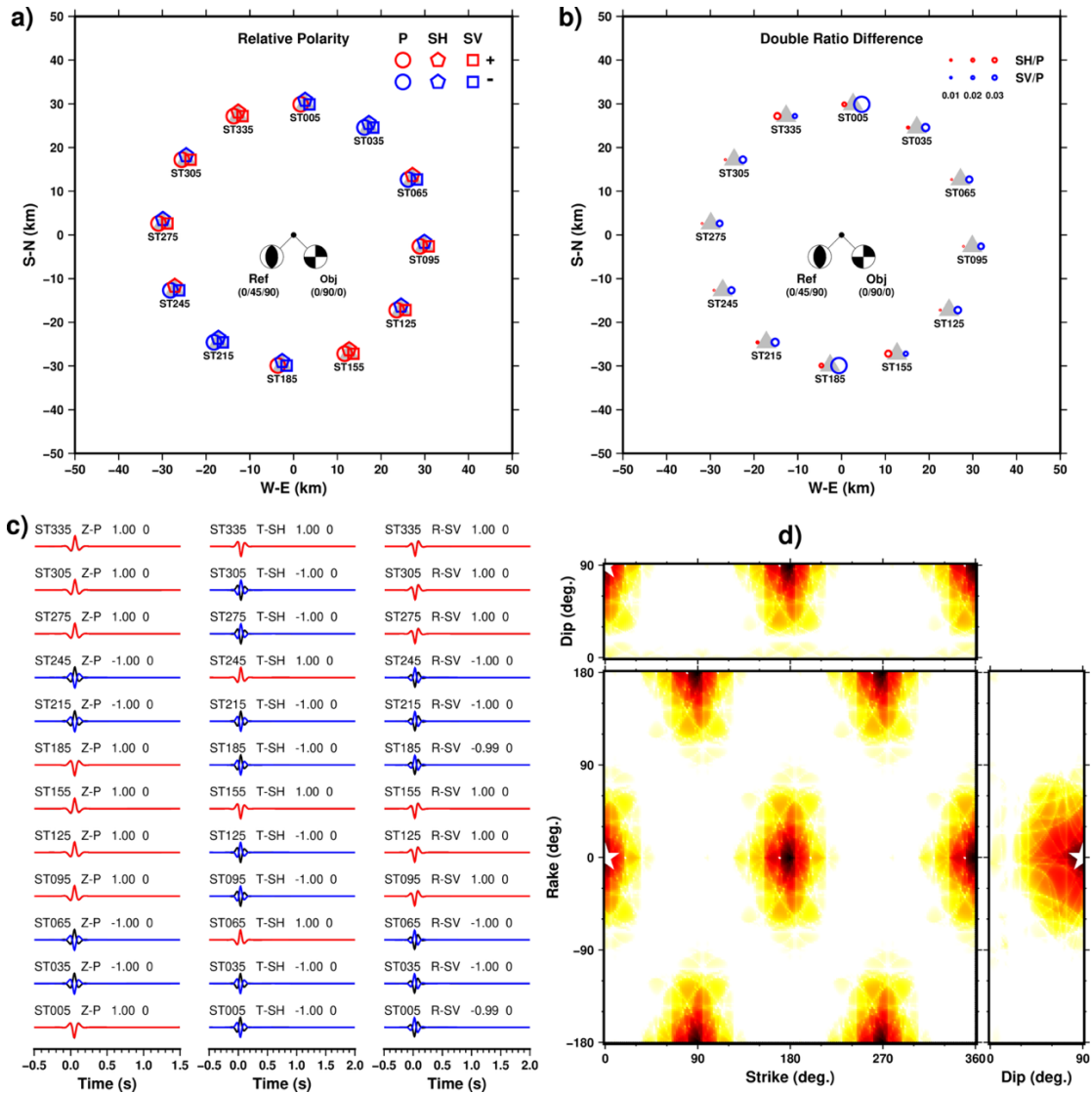

**Figure S5.** Similar to Figure 2 in the main text, except that the target event has a vertical strike-slip focal mechanism.

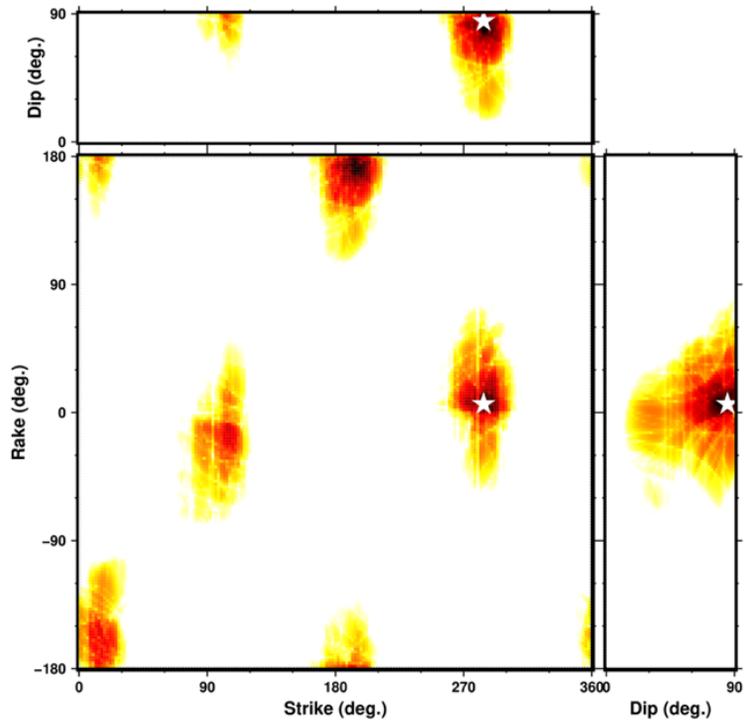

**Figure S6.** Misfit distribution in strike–dip–rake parameter space for the antisimilar earthquake pair shown in Figure 4 of the main text. The white star marks the optimal focal mechanism corresponding to the minimum misfit.

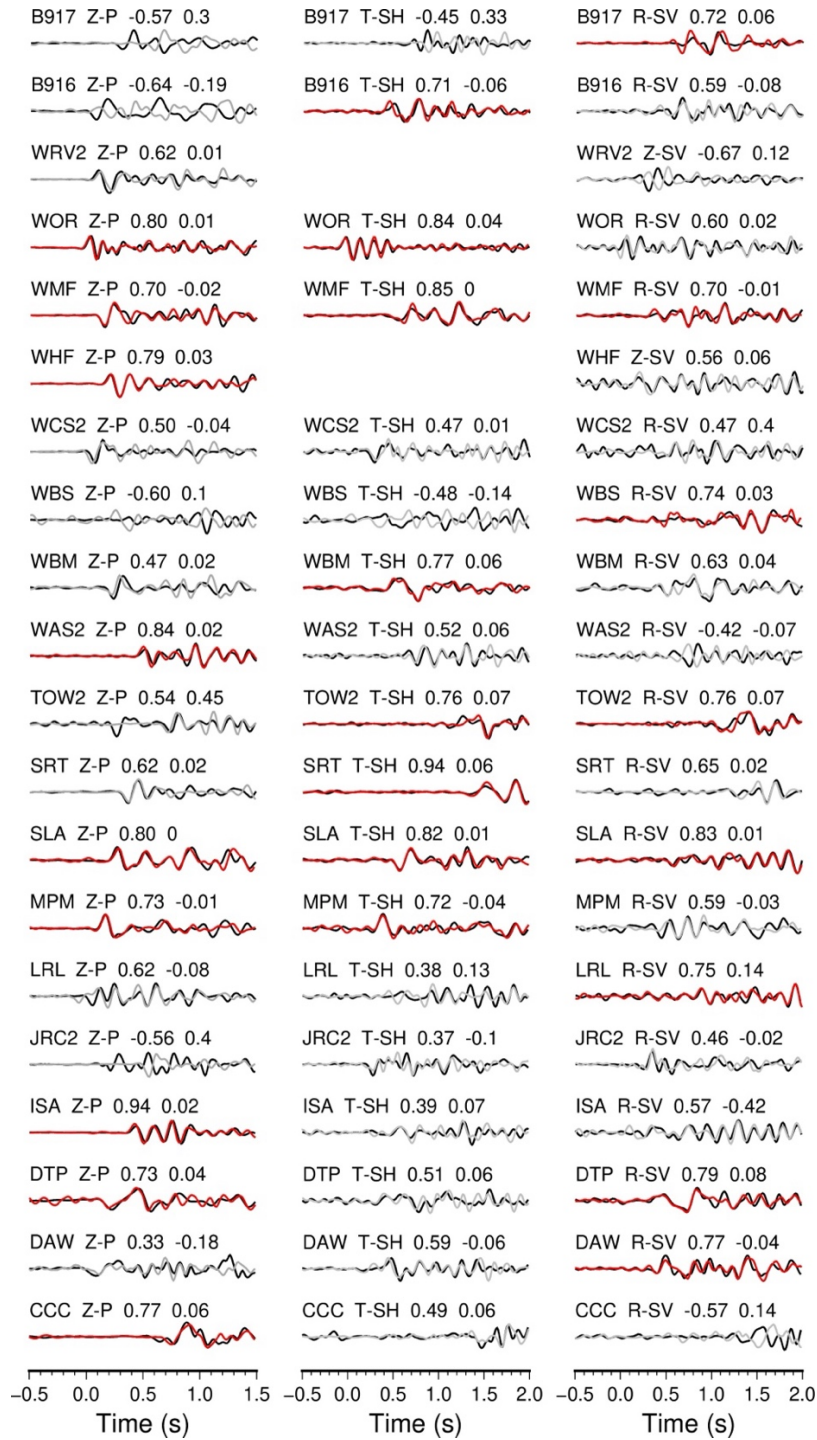

**Figure S7.** Waveform comparisons of the P, SH, and SV phases on the Z, T, and R components, respectively, for the pair of events shown in Figure 7a–b of the main text. Colored traces correspond to the target event: red indicates phases expected to be consistent with the reference event, blue indicates phases expected to have opposite polarity (none are present here), and purple indicates unexpected polarity mismatches (none are present here). The texts above each trace indicate the station name, component–phase label, the maximum normalized CC within the window, and the corresponding phase shift associated with that maximum normalized CC.

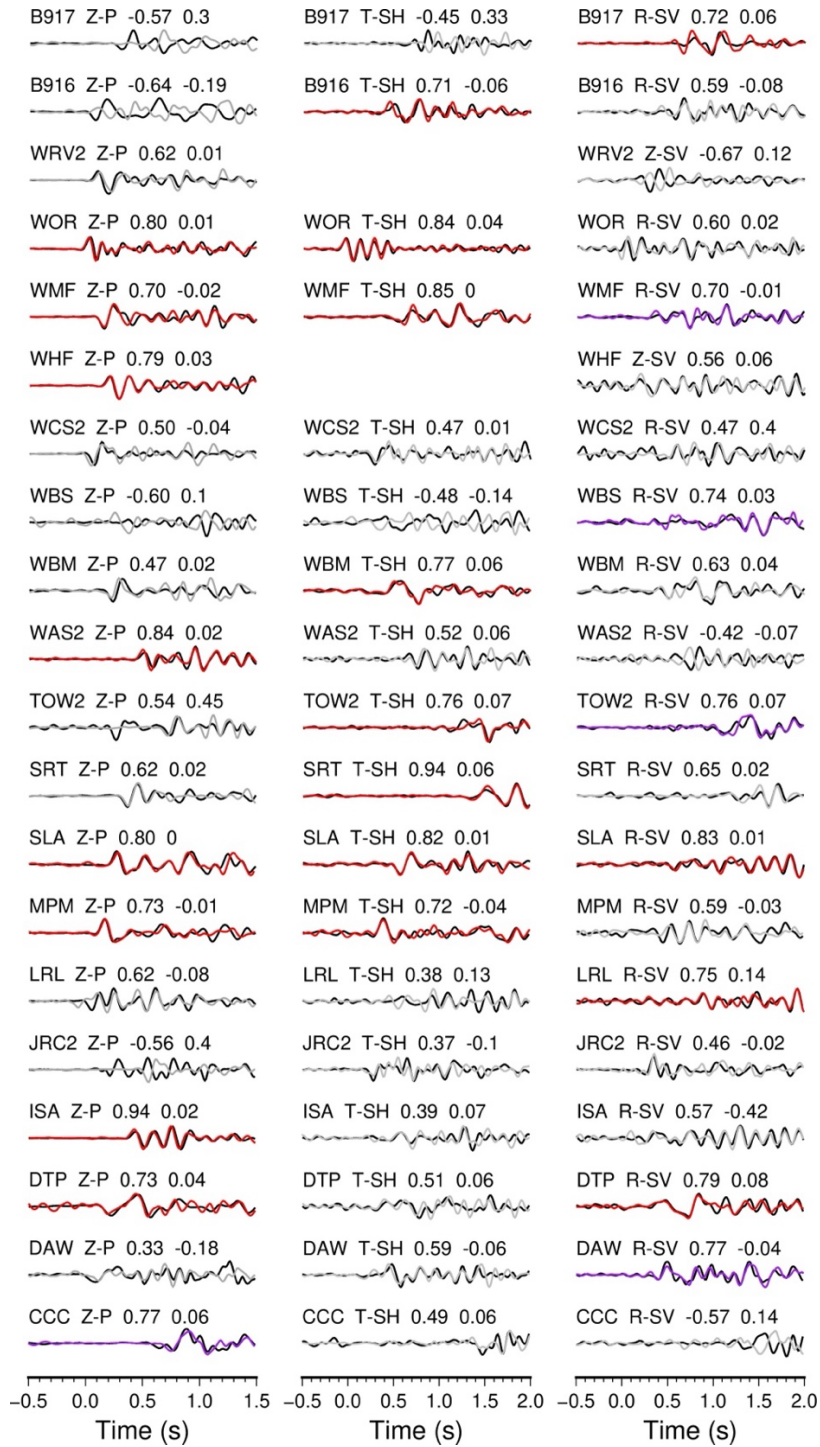

**Figure S8.** Waveform comparisons of the P, SH, and SV phases on the Z, T, and R components, respectively, for the pair of events shown in Figure 7c–d of the main text. Colored traces correspond to the target event: red indicates phases expected to be consistent with the reference event, blue indicates phases expected to have opposite polarity (none are present here), and purple indicates unexpected polarity mismatches. The texts above each trace indicate the station name, component–phase label, the maximum normalized CC within the window, and the corresponding phase shift associated with that maximum normalized CC.

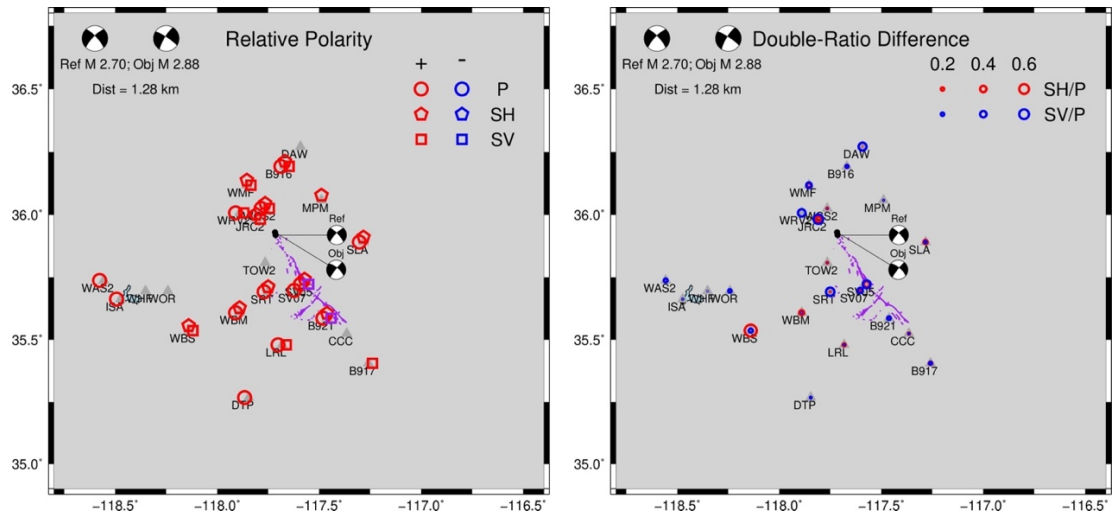

**Figure S9.** FocMecDR solving the M 2.88 event using a nearby M 2.7 reference event. (a) Relative polarities of P (circles), SH (pentagons), and SV (squares) for the event pair, with red indicating expected consistent polarities, blue indicating expected opposite polarities (none present), and purple indicating mismatches. (b) Logarithmic double-ratio misfits for SH/P (red circles) and SV/P (blue circles), with circle size proportional to the misfit at each station.

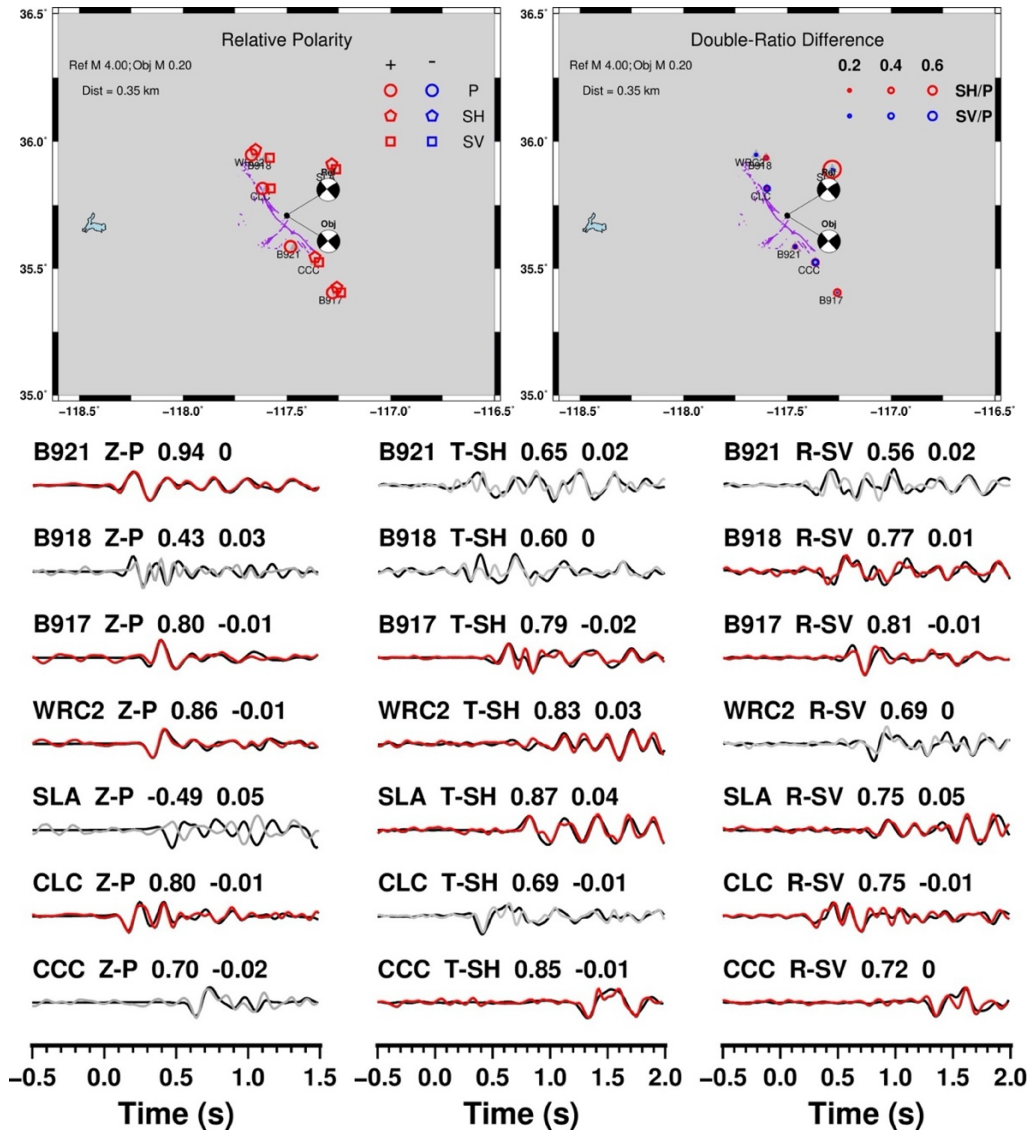

**Figure S10.** FocMecDR solving an M 0.2 foreshock event using the largest M 4.0 foreshock as the reference event. Left top panel: Relative polarities of P (circles), SH (pentagons), and SV (squares) for the event pair, with red indicating the expected consistent polarities, blue indicating the expected opposite polarities (not present here), and purple indicating unexpected polarity mismatches (not present here). Right top panel: Logarithmic double-ratio misfits for SH/P (red circles) and SV/P (blue circles), with circle size proportional to the misfit at each station. Bottom panel: Waveform comparisons of the P, SH, and SV phases on the Z, T, and R components, respectively. Colored traces show the target event: red indicates phases expected to be consistent with the reference event, blue indicates phases expected to be opposite in polarity (not present here), and purple indicates unexpected polarity mismatches (not present here). The texts above each trace represent the station name, component-phase name, the maximum normalized CC within the window, and the corresponding phase shift associated with that maximum normalized CC. The focal mechanism for the target event is shown together with its location, along with the reference event and its focal mechanism. In panels (a) and (b), mapped surface ruptures from Ponti et al. (2020) are shown in purple.